\documentclass{ws-procs961x669}           
\begin{document}
\title{Observational constraints on nonlinear matter extensions of general relativity}

\author{E.-A. Kolonia}
\address{Department of Physics, University of Patras,\\
26504 Patras, Greece}

\author{C. J. A. P. Martins$^*$}
\address{Centro de Astrof\'{\i}sica da Universidade do Porto, and\\
Instituto de Astrof\'{\i}sica e Ci\^encias do Espa\c co, Universidade do Porto,\\
Rua das Estrelas, 4150-762 Porto, Portugal\\
$^*$E-mail: Carlos.Martins@astro.up.pt}

\begin{abstract}
We present a phenomenological analysis of current observational constraints on classes of FLRW cosmological models in which the matter side of Einstein's equations includes, in addition to the canonical term, a term proportional to some function of the energy-momentum tensor ($T^2=T_{\alpha\beta}T^{\alpha\beta}=\rho^2+3p^2$), or of its trace ($T=\rho-3p$). Qualitatively, one may think of these models as extensions of general relativity with a nonlinear matter Lagrangian. As such they are somewhat different from the usual dynamical dark energy or modified gravity models: in the former class of models one adds further dynamical degrees of freedom to the Lagrangian (often in the form of scalar fields), while in the latter the gravitational part of the Lagrangian is changed. We study both of these models under two different scenarios: (1) as phenomenological two-parameter or three-parameter extensions of the standard $\Lambda$CDM, in which case the model still has a cosmological constant but the nonlinear matter Lagrangian leads to additional terms in Einstein's equations, which cosmological observations tightly constrain, and (2) as alternatives to $\Lambda$CDM, where there is no cosmological constant, and the nonlinear matter term would have to provide the acceleration (which would be somewhat closer in spirit to the usual modified gravity models). A comparative analysis of the observational constraints obtained in the various cases provides some insight on the level of robustness of the $\Lambda$ model and on the parameter space still available for phenomenological alternatives.
\end{abstract}

\keywords{Cosmology; Dark energy; Modified gravity; Observational constraints.}

\bodymatter

\section{Introduction}

The observational evidence for the acceleration of the universe shows that our canonical theories of cosmology and particle physics are at least incomplete, and possibly incorrect. Mapping the dark side of the universe, in order to ascertain the physical mechanism behind this acceleration, in a compelling observational task for current and future facilities. The CosmoESPRESSO team uses the universe as a laboratory to address, with precision spectroscopy and other observational, computational and theoretical tools, this and other grand-challenge questions. In what follows we highlight recent contributions of the CosmoESPRESSO team to this fundamental quest, pertaining to dark energy phenomenology.

There has been some recent interest in the so-called energy-momentum-squared gravity models \cite{Roshan}, where the matter part of Einstein's equations is modified by the addition of a term proportional to $T^2\equiv T_{\mu\nu}T^{\mu\nu}$, where $T_{\mu\nu}$ is the energy-momentum tensor. This we later extended \cite{Board,Akarsu} to the more generic form $(T^2)^n$, dubbed energy-momentum-powered gravity. Reference \citenum{Faria} provided low redshift constraints on these models, using in particular the Pantheon Type Ia supernova compilation \cite{Riess} and a compilation of 38 Hubble parameter measurements \cite{Farooq}.

In practical terms, we may think of these models as extensions to the canonical $\Lambda$CDM, in which case the model still has a cosmological constant but the nonlinear matter Lagrangian leads to additional terms in Einstein's equations, and cosmological observations can constrain these additional model parameters. Typically there are two such additional parameters: the power $n$ of the nonlinear part of the Lagrangian and a further parameter (to be defined below) quantifying the contribution of this term to the energy budget of the universe.

Alternatively, we may ask whether a suitably chosen nonlinear Lagrangian can reproduce the low redshift acceleration of the universe in a model which at low redshift only contains matter (plus a subdominant amount of radiation) but no true cosmological constant. In principle such a scenario is conceivable \cite{Roshan}. It is also somewhat closer in spirit to the usual modified gravity models---with the caveat that, as previously mentioned, in the latter models the modification occurs in the gravitational part of the Lagrangian and not in the matter part.

The analysis of Ref. \citenum{Faria} found that these models do not solve the cosmological constant problem \textit{per se}, but they can phenomenologically lead to a recent accelerating universe without a cosmological constant at the cost of having preferred values of the cosmological parameters that are somewhat different from the standard $\Lambda$CDM ones. Here we revisit and update these constraints, and also provide new constraints on models where the new terms depend on the trace of the energy-momentum tensor, $T=\rho-3p$. As before, we can consider both the scenario with a cosmological constant (in which case the model is an extension of $\Lambda$CDM) and the scenario without a cosmological constant (in which case we can check whether such models can accelerate at all). These have been qualitatively studied in the literature \cite{Godani,Velten}, and in what follows we provided quantitative constraints on them. We note that in all the analysis that follows the Hubble constant is analytically marginalized as discussed in Ref. \citenum{Anagnostopoulos}.

\section{Energy-momentum-powered models}

The general action for these models is  \cite{Board}
\begin{equation}
S=\frac{1}{2\kappa}\int\left[R+\eta (T^2)^n-2\Lambda\right]d^4x + S_{matter}\,,
\end{equation}
where $\kappa=8\pi G$, and $\eta$ is a constant quantifying the contribution of the $T^2$-dependent term. In a flat Friedmann-Lemaitre-Robertson-Walker universe and assuming a perfect fluid we have $T^2=\rho^2+3p^2$ and the generalized Friedmann and Raychaudhuri equations and the corresponding continuity equation can be written
\begin{eqnarray}
3\left(\frac{\dot a}{a}\right)^2&=&\Lambda+\kappa\rho+\eta(\rho^2+3p^2)^{n-1}\left[\left(n-\frac{1}{2}\right)(\rho^2+3p^2)+4np\rho\right]\\
6\frac{\ddot a}{a}&=&2\Lambda-\kappa(\rho+3p)-\eta(\rho^2+3p^2)^{n-1}\left[(n+1)(\rho^2+3p^2)+4np\rho\right]\\
{\dot\rho}&=&-3\frac{\dot a}{a}(\rho+p)\frac{\kappa\rho+n\eta\rho(\rho+3p)(\rho^2+3p^2)^{n-1}}{\kappa\rho+2n\eta(\rho^2+3p^2)^{n-1}\left[\left(n-\frac{1}{2}\right)(\rho^2+3p^2)+4np\rho\right]}.
\end{eqnarray}
As usual, the Bianchi identity implies that only two of these equations are independent.

If we consider the low redshift limit of these models, further assuming that the universe is composed of matter and possibly also a cosmological constant, we can simplify the Einstein equations to
\begin{eqnarray}
3\left(\frac{\dot a}{a}\right)^2&=&\Lambda+\kappa\rho+\left(n-\frac{1}{2}\right)\eta \rho^{2n}\\
6\frac{\ddot a}{a}&=&2\Lambda-\kappa\rho-(n+1)\eta\rho^{2n}\\
{\dot\rho}&=&-3\frac{\dot a}{a}\rho\frac{\kappa+n\eta\rho^{2n-1}}{\kappa+(2n-1)n\eta\rho^{2n-1}}\,.
\end{eqnarray}
Broadly speaking, we inspection of the equations leads to the expectation that $n<1/2$ may be interesting at late times. In general these equations need to be solved numerically. However, there are three particular cases for which analytic solutions can be found (at least approximate, low redshift solutions), corresponding to the values $n=1$, $n=1/2$ and $n=0$ (the latter actually corresponds to the $\Lambda$CDM case). These have been studied in the literature, in a general mathematical context \cite{Roshan,Board,Early}, and also observationally constrained in Ref. \citenum{Faria}.

Generically we can treat $n$ as a further model parameter, to be constrained by observations. In order to do this we define a dimensionless cosmological density $r$, via $\rho = r \rho_0$, where $\rho_0$ is the present day density, as well as a generic parameter
\begin{equation}
Q=\frac{\eta}{\kappa}\rho_0^{2n-1}\,.
\end{equation}
With these assumptions, and keeping for the time being the matter assumption, the continuity equation expressed in terms of redshift has the form
\begin{equation}
\frac{dr}{dz}=\frac{3r}{1+z}\times\frac{1+nQr^{2n-1}}{1+(2n-1)nQr^{2n-1}}\,;
\end{equation}
\begin{equation}
E^2(z)=\frac{H^2(z)}{H_0^2}=\Omega_\Lambda+\Omega_Mr+\left(n-\frac{1}{2}\right)Q\Omega_Mr^{2n}\,,
\end{equation}
where, since we must have $E(0)=1$, the model parameters are related by the requirement that $\Omega_\Lambda=1-\Omega_M[1+(n-1/2)Q]$, and therefore the Friedmann can be recast in the two alternative forms
\begin{eqnarray}
E^2(z)&=&\Omega_\Lambda+\Omega_Mr+(1-\Omega_M-\Omega_\Lambda)r^{2n} \label{friednoq} \\
E^2(z)&=&1+\Omega_M(r-1)+\left(n-\frac{1}{2}\right)Q\Omega_M(r^{2n}-1)\,,
\end{eqnarray}
the first one is generic, while the second applies only if $\Omega_\Lambda \neq0$. On the other hand, if $\Omega_\Lambda=0$ we can also use the flatness assumption to eliminate $Q$ in the continuity equation, writing it as
\begin{equation}
\frac{dr}{dz}=\frac{3r}{1+z}\times\frac{(2n-1)\Omega_M+2n(1-\Omega_M)r^{2n-1}}{(2n-1)[\Omega_M+2n(1-\Omega_M)r^{2n-1}]}\,.    
\end{equation}
This shows that in a phenomenological sense these models could explain the recent acceleration of the universe without invoking a cosmological constant but relying instead on the nonlinearities of the matter Lagrangian in a matter-only universe with $n=0$. However, we note that even if these models can lead to accelerating universes without a cosmological constant, this is not per se sufficient to solve the 'old' cosmological constant problem of why it should be zero. For $n$ close to but not equal two zero, two things happen: the (formerly) constant term  in the Friedmann equation becomes slowly varying, and the continuity equation implies that the matter density does not behave exactly as $r\propto (1+z)^3$. In what follows we discuss the extent to which deviations from the $n=0$ case are observationally allowed, and also the most general parameter space where a standard cosmological constant is also allowed.

We can also consider a generalization: instead of considering a universe with a matter fluid, we can assume that this fluid has a constant equation of state $w=p/\rho=const.$ (with the matter case corresponding to $w=0$). In this case the continuity equation becomes
\begin{equation}
\frac{dr}{dz}=\frac{3r}{1+z}(1+w)\times\frac{1+nQf_1(n,w)r^{2n-1}}{1+2nQf_2(n,w)r^{2n-1}}\,,
\end{equation}
where for convenience we defined
\begin{eqnarray}
f_1(n,w)&=&(1+3w)(1+3w^2)^{n-1}\\
f_2(n,w)&=&(1+3w^2)^{n-1}\left[\left(n-\frac{1}{2}\right)(1+3w^2)+4nw\right]\,.
\end{eqnarray}
In this case the Friedmann equation can be written
\begin{equation}
E^2(z)=\Omega_\Lambda+\Omega_Mr+f_2(n,w)Q\Omega_Mr^{2n}\,,
\end{equation}
together with the consistency relation $\Omega_\Lambda=1-\Omega_M[1+f_2Q]$. It follows that we can also re-write it as
\begin{eqnarray}
E^2(z)&=&\Omega_\Lambda+\Omega_Mr+(1-\Omega_M-\Omega_\Lambda)r^{2n}\\
E^2(z)&=&1+\Omega_M(r-1)+f_2(n,w)Q\Omega_M(r^{2n}-1)\,,
\end{eqnarray}
where again the first is generic---and indeed identical to Eq. (\ref{friednoq}), although the redshift dependence of $r$ will now be different---while the second holds for $\Omega_\Lambda\neq0$. Here, if $\Omega_\Lambda=0$ the continuity equation can also be written in a way that eliminates the parameter $Q$,
\begin{equation}
\frac{dr}{dz}=\frac{3r}{1+z}(1+w)\times\frac{\Omega_Mf_2+n(1-\Omega_M)f_1r^{2n-1}}{f_2[\Omega_M+2n(1-\Omega_M)r^{2n-1}]}\,.
\end{equation}

\begin{table}
\tbl{One sigma posterior likelihoods on the power $n$, the matter density $\Omega_M$ and the constant equation of state $w$ (when applicable) for various flat energy-momentum-powered models containing matter, with or without a cosmological constant. The last column lists the reduced chi-square for each best-fit model.}
{\begin{tabular}{lcccc}
\toprule
Model assumptions & $\Omega_M$ & $n$  & $w$ & $\chi^2_\nu$ \\
\colrule
$\Omega_\Lambda=0$, $w=0$ & $0.39\pm0.08$ & $0.04\pm0.04$ & N/A & 0.64\\
$\Omega_\Lambda\neq0$, $w=0$ & $0.29^{+0.05}_{-0.03}$ & Unconstrained & N/A & 0.64 \\
$\Omega_\Lambda=0$, $w=const.$  & $0.28^{+0.12}_{-0.10}$ & $-0.08^{+0.06}_{-0.02}$& $-0.11^{+0.07}_{-0.04}$ & 0.62 \\
\botrule
\end{tabular}}
\begin{tabnote}
The specific assumptions for each case are described in the main text. The constraints come from the combination of the Pantheon supernova data and Hubble parameter measurements.
\end{tabnote}\label{table1}
\end{table}

\section{Constraints on energy-momentum-powered models}

In what follows we briefly summarize constraints on the generic energy-momentum-powered models, revising and updating the analysis in Ref. \citenum{Faria}. We carry out a standard likelihood analysis, using the datasets already mentioned in the introduction, and separately considering three different theoretical scenarios within this class of models. An overview of the results can be found in Table \ref{table1}.

\begin{figure}
\begin{center}
\includegraphics[width=\columnwidth,keepaspectratio]{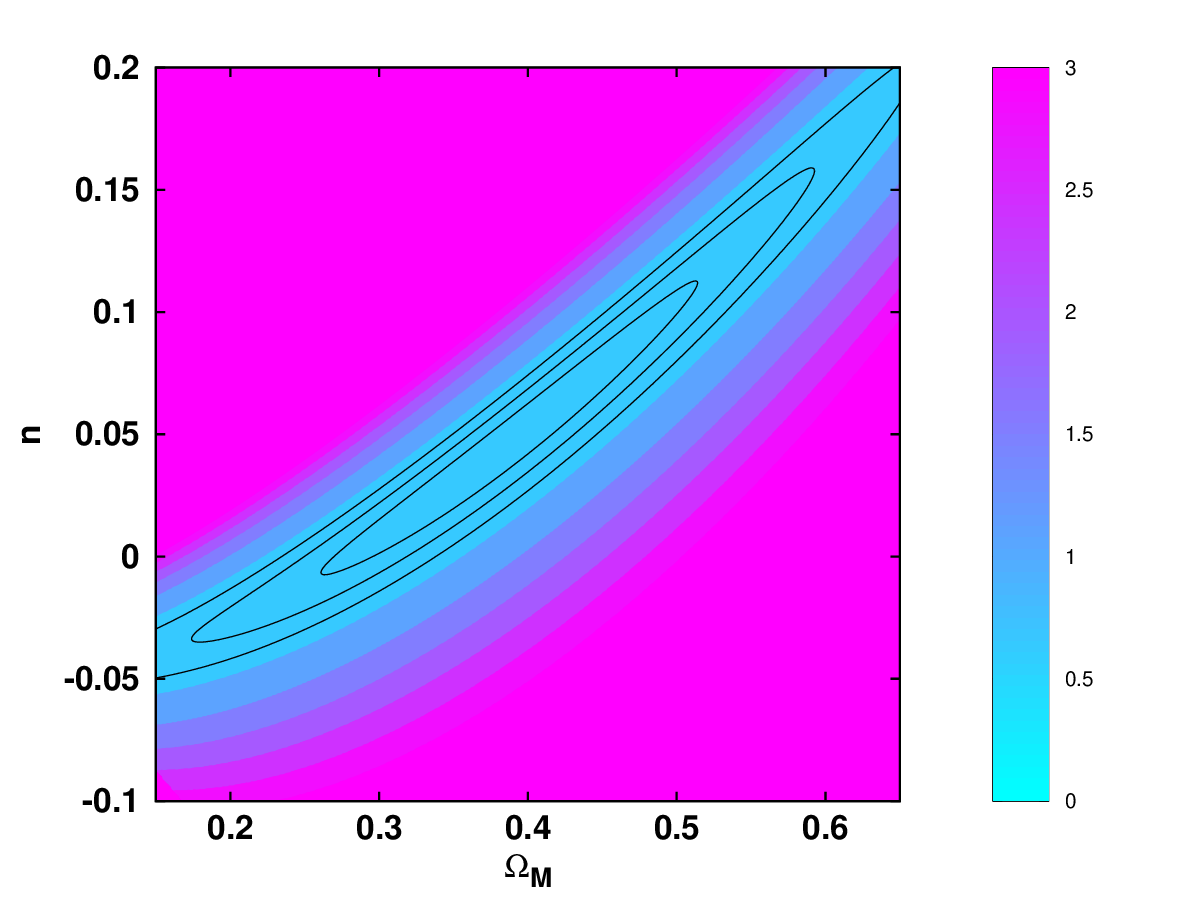}
\includegraphics[width=\columnwidth,keepaspectratio]{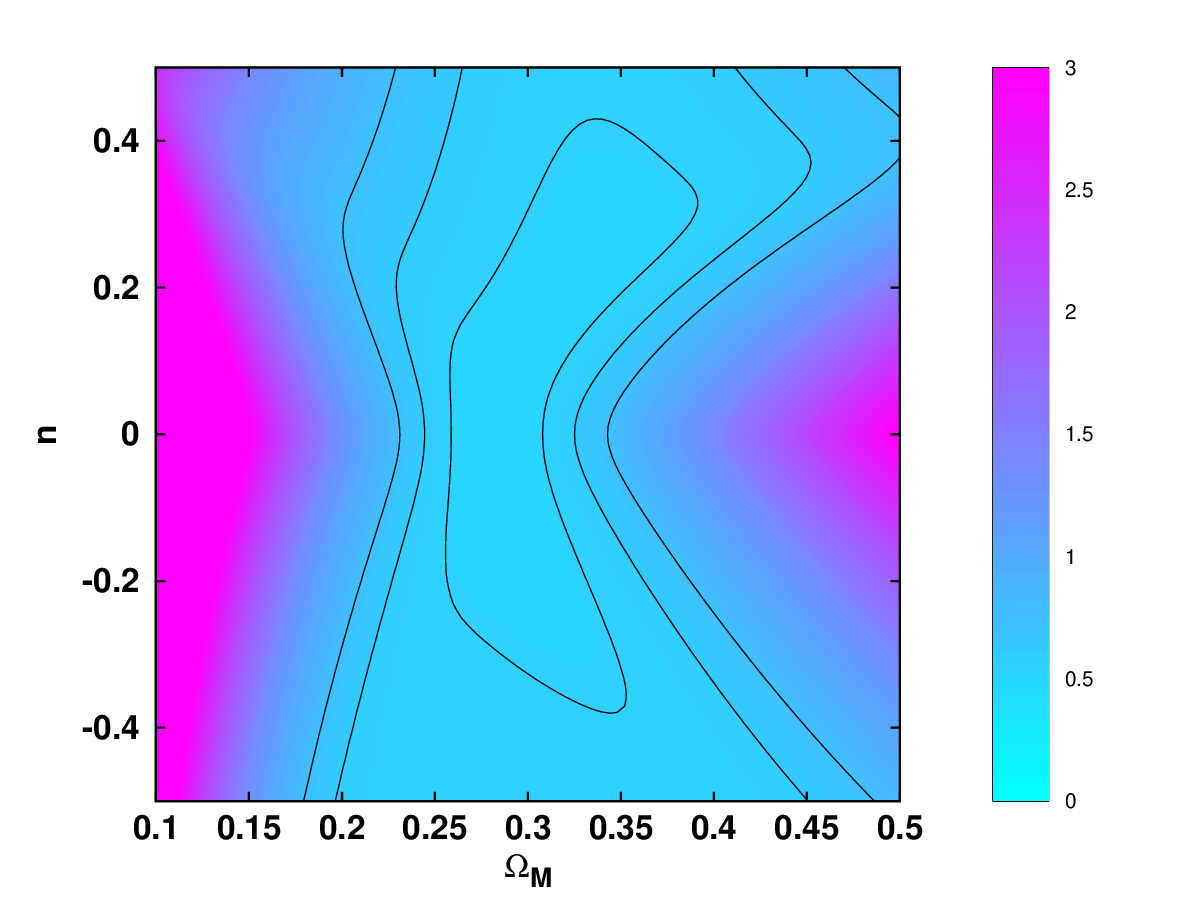}
\end{center}
\caption{Constraints on the $n$--$\Omega_M$ parameter space for flat universes with $w=0$. Top and bottom panels correspond to the  $\Omega_\Lambda=0$ and $\Omega_\Lambda\neq0$ cases, respectively. The black solid curves show the one, two, and three sigma confidence levels, while the color map depicts the reduced chi-square.}
\label{figure1}
\end{figure}

The constraints for the $\Omega_\Lambda=0$ matter case are summarized in the top panel pf Fig. \ref{figure1} and also in the first row of Table \ref{table1}. As expected given the form of the Friedmann and continuity equations, there is a clear degeneracy between the two parameters. The best-fit values are about one standard deviation away from the canonical values $n=0$ and $\Omega_M\sim0.3$, and a non-zero $n\sim0.04$ and a slightly higher matter density are preferred. However, at the two sigma level the results are consistent with $\Lambda$CDM; one should also bear in mind that the $n=0$ does correspond to the $\Lambda$CDM case. 

Constraints on the $\Omega_\Lambda\neq0$ matter case are shown in the bottom panel of Fig. \ref{figure1} and also in the second row of Table \ref{table1}. Here there is a strong degeneracy between $Q$ and $n$, both of which are unconstrained. On the other hand, the matter density is still well constrained (indeed, the constraint is tighter than in the case without a cosmological constant) and fully consistent with the canonical $\Lambda$CDM value.

\begin{figure}
\begin{center}
\includegraphics[width=\columnwidth,keepaspectratio]{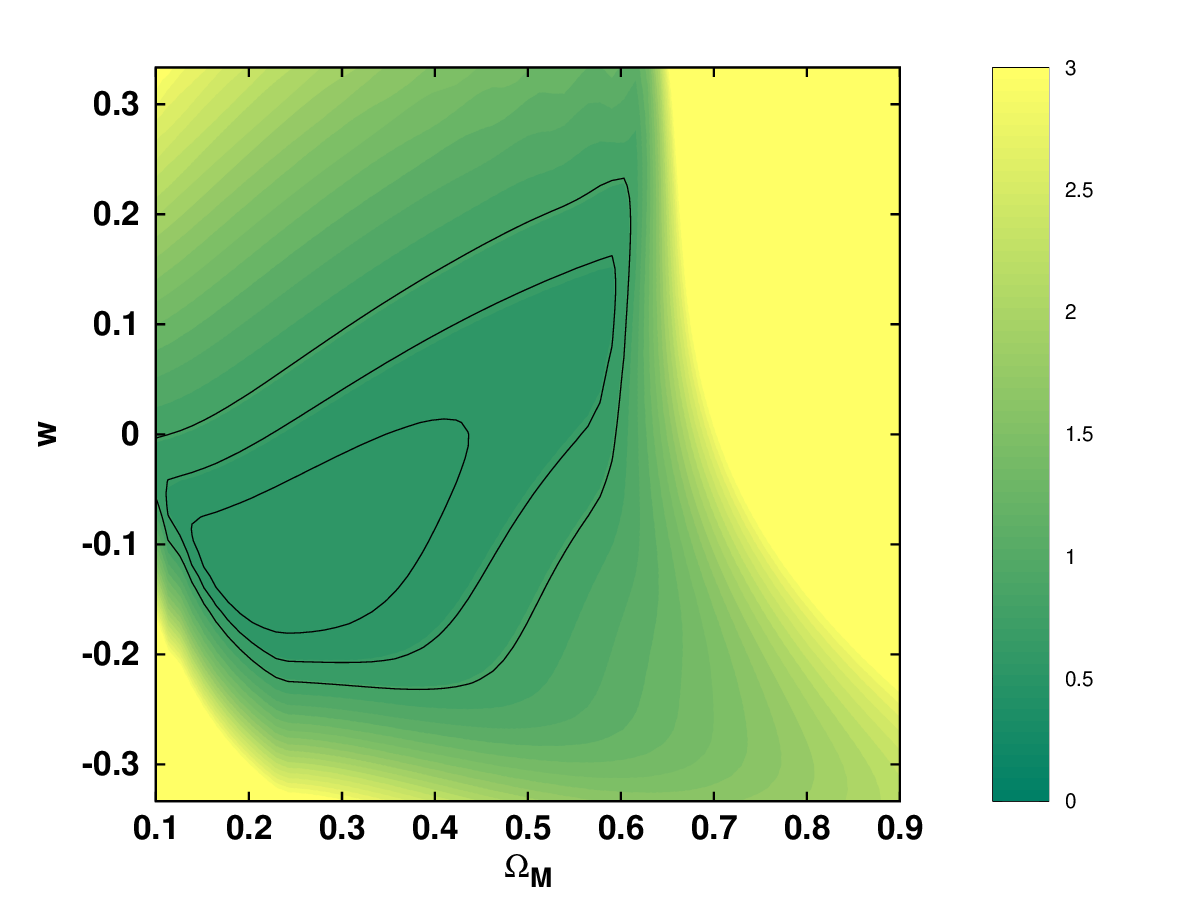}
\includegraphics[width=\columnwidth,keepaspectratio]{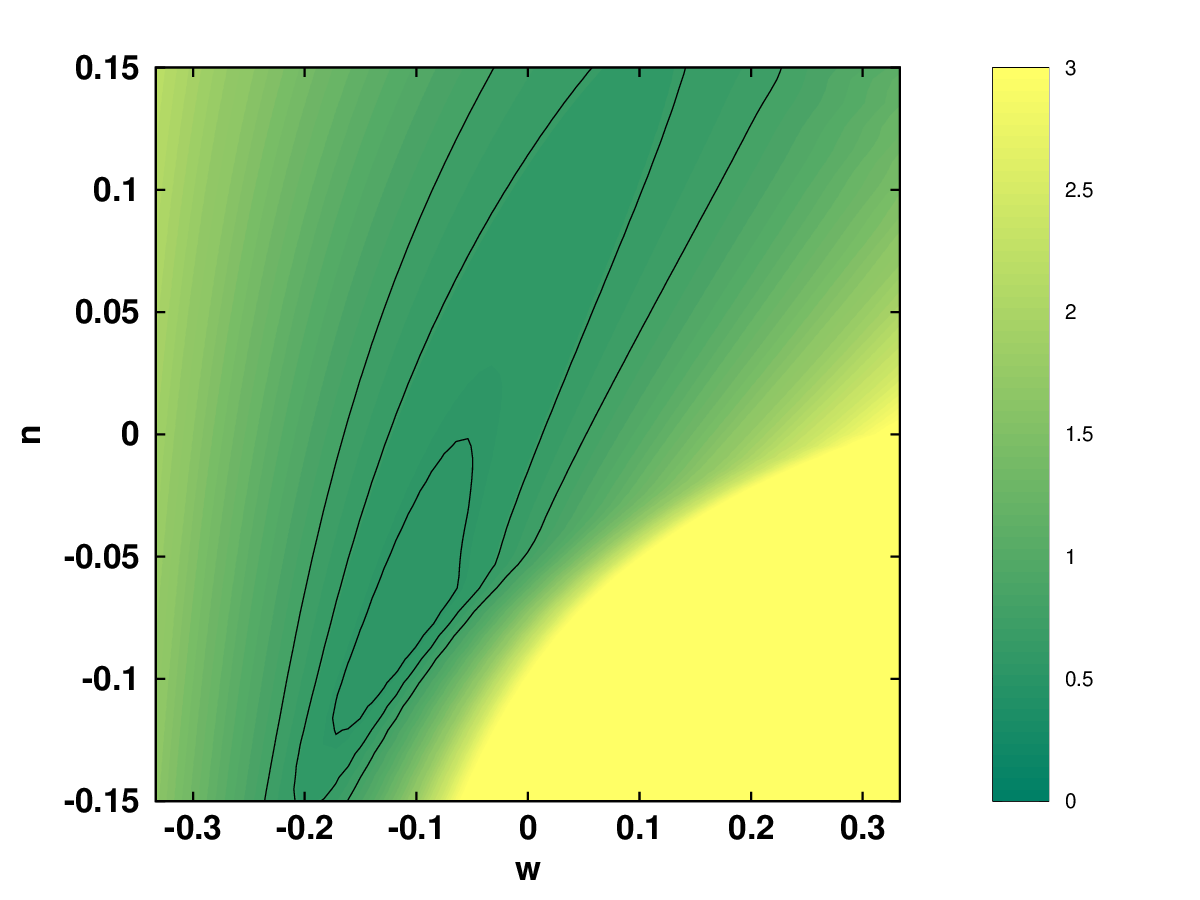}
\end{center}
\caption{Constraints on the $n$--$\Omega_M$--$w$ parameter space for flat universes with $\Omega_\Lambda=0$. The black solid curves show the one, two, and three sigma confidence levels, while the color maps depict the reduced chi-square.}
\label{figure2}
\end{figure}

Finally, constraints on the $\Omega_\Lambda=0$ case while allowing for a constant equation of state, $w=const$ are in Fig. \ref{figure2} and also in the final row of Table \ref{table1}. In this case one can constrain the three model parameters. The preferred value of the matter density is again consistent with the canonical $\Lambda$CDM value, while there is a mild preference (less than two standard deviations) for negative values on the exponent $n$ and the equation of state parameter $w$.

Overall, we note that the best-fit value for the matter density is compatible, within the uncertainties, with the standard one, and there is no significant evidence for deviations from $\Lambda$CDM. On the other hand, it is worthy of note that the values of the reduced chi-square for all the best-fit models is significantly below unity, so the models clearly overfit the low redshift data that we are considering.

\section{A simple $f(R,T)$ model}

We will now explore the modified gravity model recently discussed in Ref. \citenum{Godani}. This is actually one case of a larger set of $f(R,T)$ models, to be discussed elsewhere.  A class of modified gravity models now dubbed $f(R)$ gravity,  $R$ being the scalar curvature, was first considered in Ref. \citenum{Buchdahl}, but these models are subject to tight cosmological constraints \cite{Clifton}. A phenomenologically  broader (if physically less well motivated) class is that of the so-called $f(R,T)$ models, where $T$ is the trace of the stress energy tensor. A particular subclass of these models has separable function, $f(R,T)=R+f_2(T)$,. These models have been the subject of several mathematical studies but so far they have not been put through a detailed comparison with cosmological observations, with the exception of the recent qualitative analysis of Ref. \citenum{Velten}.

Qualitatively, the main difference is that here the new terms depend on the trace of the energy-momentum tensor $T=\rho-3p$, while in the model considered in the previous sections they depended on $T^2\equiv T_{\mu\nu}T^{\mu\nu}=\rho^2+3p^2$. The procedure for studying the two models should otherwise be similar.

This model \cite{Godani}, also previously considered in Ref. \citenum{Velten}, has the action
\begin{equation}
S=\frac{1}{2\kappa}\int\left[R+\xi \sqrt{T}-2\Lambda\right]d^4x + S_{matter}\,.
\end{equation}
In a flat FLRW universe the Friedmann and Raychaudhuri equations now have the following form\footnote{We note that there is a minus sign typo in the last term of Eq. (10) in Ref. \citenum{Godani}.}
\begin{eqnarray}
3\left(\frac{\dot a}{a}\right)^2&=&\Lambda+\kappa\rho+\xi\frac{(\rho-p)}{\sqrt{\rho-3p}}\\
6\frac{\ddot a}{a}&=&2\Lambda-\kappa(\rho+3p)+\frac{\xi}{2}\frac{(\rho-7p)}{\sqrt{\rho-3p}}\,.
\end{eqnarray}
As a simple comparison, in the $p=0$ case this model leads to a Friedmann equation of the form
\begin{equation}
3H^2=\Lambda+\kappa\rho+\xi\sqrt{\rho}\,,    
\end{equation}
while in the energy-momentum-powered model, choosing $n=1/4$, one has
\begin{equation}
3H^2=\Lambda+\kappa\rho-\frac{\eta}{4}\sqrt{\rho}\,.
\end{equation}
We note that the two Friedmann equations coincide (if one identifies $\xi=-\eta/4$), but the corresponding continuity equations differ in the two cases.

We will again assume constant equations of state ($p=w\rho$), use the standard definitions of $\Omega_M$ and $\Omega_\Lambda$ together with $\rho=r\rho_0$ and additionally define
\begin{equation}
\zeta=\frac{\xi}{2\kappa\sqrt{\rho_0}}\,.    
\end{equation}
We can then rewrite the Friedmann equation as follows
\begin{equation}
E^2(z)=\Omega_\Lambda+\Omega_Mr+2\zeta\frac{(1-w)}{\sqrt{1-3w}}\Omega_M\sqrt{r}\,.
\end{equation}
In principle there are therefore 3 free parameters, since the $E(0)=1$ condition requires that the model parameters are related by $\Omega_\Lambda=1-\Omega_M[1+2\zeta(1-w)/\sqrt{1-3w}]$. We can also rewrite it as
\begin{eqnarray}
E^2(z)&=&\Omega_\Lambda+\Omega_Mr+(1-\Omega_M-\Omega_\Lambda)\sqrt{r}\\
E^2(z)&=&1+\Omega_M(r-1)+2\zeta\frac{(1-w)}{\sqrt{1-3w}}\Omega_M(\sqrt{r}-1)\,;
\end{eqnarray}
the first of these is generic, while the second is only valid if $\Omega_\Lambda\neq0$. However, note that in general the parameters $(\zeta,w)$ still affect the continuity equation, which can be written as
\begin{equation}
\frac{dr}{dz}=\frac{3r}{1+z}(1+w)\times\frac{\sqrt{1-3w}+\zeta/\sqrt{r}}{\sqrt{1-3w}+(1-w)\zeta/\sqrt{r}}\,.    
\end{equation}
We note that the usual behaviour, $r\propto (1+z)^3$, is recovered for $\zeta=0$ and that (less trivially) this also occurs for the matter case ($w=0$) for any value of the parameter $\zeta$. As an illustration of the role of this parameter we can solve the continuity in the $\zeta\to0$ limit. One finds
\begin{equation}
r(z)=\left[\left(1+\frac{w\zeta}{\sqrt{1-3w}}\right)(1+z)^{3(1+w)/2}-\frac{w\zeta}{\sqrt{1-3w}}\right]^2\,,    
\end{equation}
which again has the appropriate limits.

\begin{table}
\tbl{One sigma posterior likelihoods on the matter density $\Omega_M$, the coupling $\zeta$ and the constant equation of state $w$ (when applicable) for various flat $\sqrt{T}$ models containing matter, with or without a cosmological constant. The last column lists the reduced chi-square for each best-fit model.}
{\begin{tabular}{lcccc}
\toprule
Model assumptions & $\Omega_M$ & $\zeta$ & $w$ & $\chi^2_\nu$ \\
\colrule
$\Omega_\Lambda=0$, $w=0$ & $0.15\pm0.02$ & $2.78\pm1.72$ & N/A & 1.80\\
$\Omega_\Lambda\neq0$, $w=0$ & $0.25^{+0.03}_{-0.02}$ & $0.23^{+0.22}_{-0.18}$ & N/A & 0.63 \\
$\Omega_\Lambda=0$, $w=const.$ & $0.24^{+0.08}_{-0.07}$ & Unconstrained & $-0.08^{+0.04}_{-0.05}$ & 1.80 \\
\botrule
\end{tabular}}
\begin{tabnote}
The specific assumptions for each case are described in the main text. The constraints come from the combination of the Pantheon supernova data and Hubble parameter measurements.
\end{tabnote}\label{table2}
\end{table}

\section{Observational constraints on the $\sqrt{T}$ model}

The model can now be constrained, and in particular we can again consider both the scenario with a cosmological constant (in which case the model is an extension of $\Lambda$CDM) and the scenario without a cosmological constant (in which case we can check whether such models can account for the recent acceleration of the universe at all). An overview of the results can be found in Table \ref{table2}.

\begin{figure}
\begin{center}
\includegraphics[width=\columnwidth,keepaspectratio]{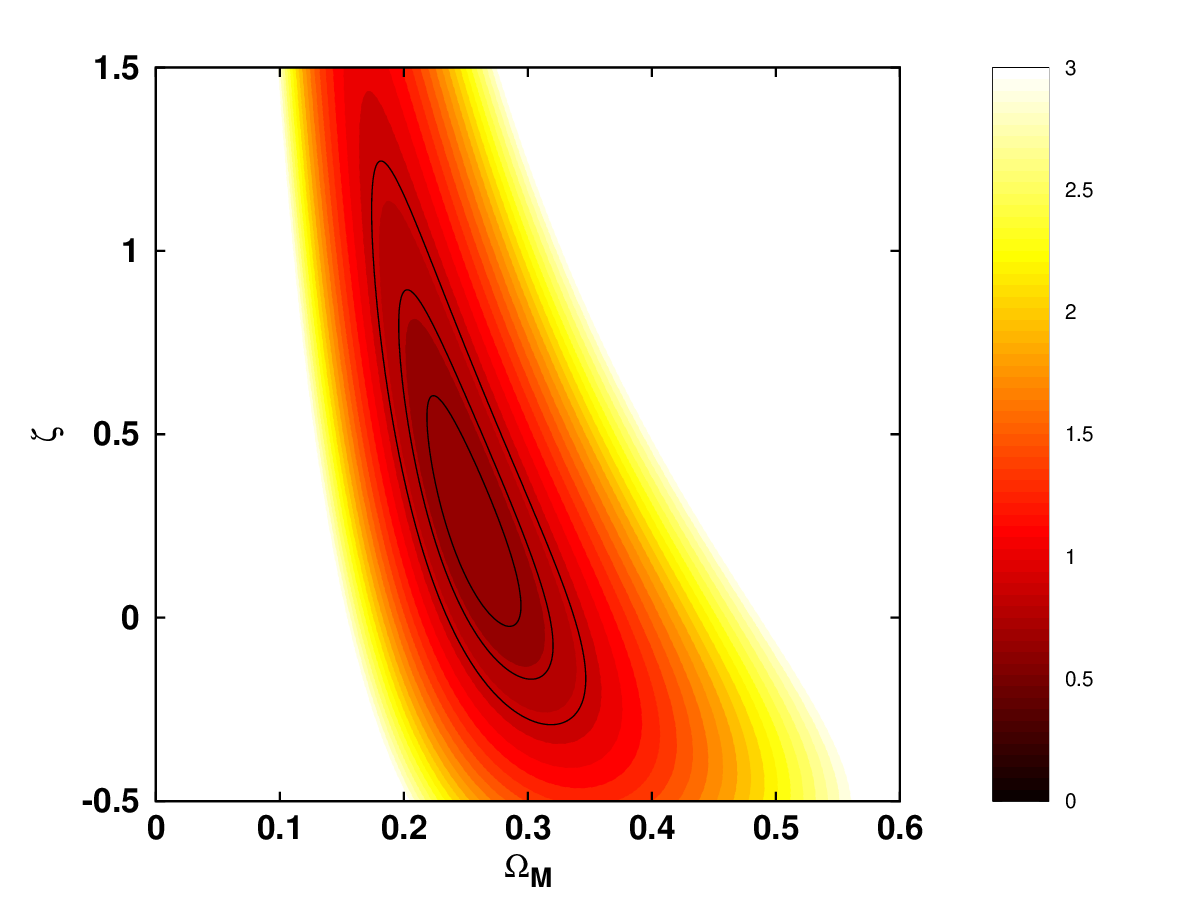}
\includegraphics[width=\columnwidth,keepaspectratio]{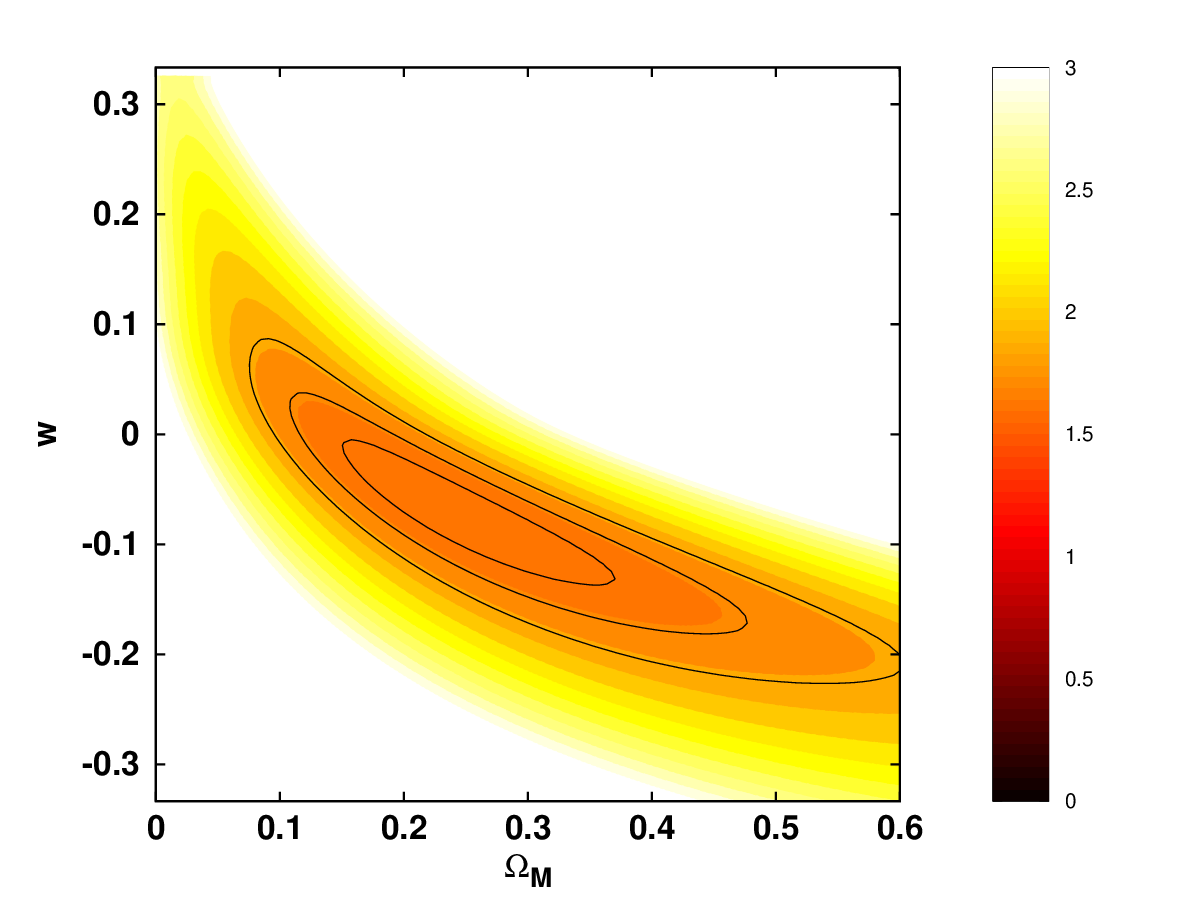}
\end{center}
\caption{Constraints on flat $\sqrt{T}$ models. The top panel shows constraints for $\Omega_\Lambda\neq0$ and $w=0$, and the bottom panel shows constraints for $\Omega_\Lambda=0$ and $w=const$. The black solid curves show the one, two, and three sigma confidence levels, and the color maps depict the reduced chi-square.}
\label{figure3}
\end{figure}

Starting with the case of $\Omega_\Lambda=0$ and $w=0$, we effectively have only one independent parameter, since the matter density and coupling parameter are related through $(1+2\zeta)\Omega_M=1$. The obtained constraints are listed in the first row of Table \ref{table2}. It is clear from the very large value of the reduced chi-square that this model, containing only matter but no cosmological constant, does not fit the data. In other words, this model can not be a true alternative to $\Lambda$CDM.

The top panel of Fig. \ref{figure3} and the second row of Table \ref{table2} summarize the constraints for the case of $\Omega_\Lambda\neq0$ and $w=0$. In this case we have two independent parameters, and the model is effectively a one parameter extension of $\Lambda$CDM. Here, as in the previously discussed case of energy-momentum-powered models, the model slightly overfits the data, and there is no statistically significant preference for a non-zero coupling parameter $\zeta$, The best fit value of the matter density is not significantly changed with respect to its value, for the same datasets, in the $\Lambda$CDM model \cite{Fernandes}.

Finally, the bottom panel of Fig. \ref{figure3} and the third row of Table \ref{table2} show the constraints for $\Omega_\Lambda=0$ and $w\neq0$, where we have three independent parameters. There are now additional degeneracies between the parameters, but the main conclusion remains the same as for the $w=0$ case: without a cosmological constant this model severely underfits the data, and therefore it is not viable as an alternative to $\Lambda$CDM.

\section{Outlook}

We have discussed  observational current low redshift background constraints on classes of FLRW cosmological models in which the matter side of Einstein's equations includes, in addition to the canonical term, either a term proportional to a function of the energy-momentum tensor ($T^2=\rho^2+3p^2$), or of its trace ($T=\rho-3p$). Both of these can be phenomenologically thought of as extensions of general relativity with a nonlinear matter Lagrangian. 

We considered both models under two different scenarios: (1) as phenomenological two-parameter or three-parameter extensions of the standard $\Lambda$CDM, in which case the model still has a cosmological constant but the nonlinear matter Lagrangian leads to additional terms in Einstein's equations, which cosmological observations tightly constrain, and (2) as alternatives to $\Lambda$CDM, where there is no cosmological constant, and the nonlinear matter term would have to provide the acceleration (which would be somewhat closer in spirit to the usual modified gravity models).

Overall, our analysis suggests that the $\Lambda$CDM paradigm is fairly robust or, pragmatically, it is a good phenomenological approximation to a still unknown more fundamental model. In other words, if there is no true cosmological constant, the alternative mechanism must effectively be like one, at least at low redshifts. On the other hand, for parametric extensions of $\Lambda$CDM, subdominant (ca. $10\%$ level) contributions are allowed by the low redshift background cosmology data that we have considered. These constraints can of course be tightened by including additional datasets, such as that from cosmic microwave background observations. Our work can be extended to broader classes of $f(R,T)$ models, a discussion of which is left for a subsequent publication.

\section*{Acknowledgments}

This work was financed by FEDER---Fundo Europeu de Desenvolvimento Regional funds through the COMPETE 2020---Operational Programme for Competitiveness and Internationalisation (POCI), and by Portuguese funds through FCT - Funda\c c\~ao para a Ci\^encia e a Tecnologia in the framework of the project POCI-01-0145-FEDER-028987 and PTDC/FIS-AST/28987/2017. The project that led to this work was started during AstroCamp 2020.

\eject

\bibliographystyle{ws-procs961x669}
\bibliography{martinsnonlinear}

\end{document}